\documentclass[aps,twocolumn,a4paper,superscriptaddress,amsmath,amssymb,amsfonts]{revtex4}
\usepackage{graphicx} 
\usepackage{graphicx}
\usepackage{graphics}
\usepackage{amsmath}
\usepackage{amsthm} %
\usepackage{amssymb}%
\usepackage{amsfonts}
\usepackage{amssymb}
\usepackage{epstopdf}
\usepackage{makeidx}
\usepackage{epsfig}
\usepackage{amsfonts}
\usepackage{bm}
\usepackage{color}
\usepackage{xcolor}
\usepackage{bbold}
\usepackage{afterpage}
\usepackage{empheq}
\usepackage{hyperref}
\usepackage{booktabs}
\usepackage{subfigure}

\newcommand{\om}{\iffalse}

\newcommand{\ba}{\arraycolsep 0.3ex \begin{array}{rl}}
\newcommand{\ea}{\end{array}}
\newcommand{\bc}{\begin{cases}}
\newcommand{\ec}{\end{cases}}
\hypersetup{colorlinks,citecolor=blue,filecolor=blue,linkcolor=blue,urlcolor=blue }

\date{\today}

\begin{document}

\title{Room-Temperature Disorder-Driven Nonlinear Transport in Topological Materials}

\author{Rhonald Burgos Atencia}
\affiliation{School of Physics, The University of New South Wales, Sydney 2052, Australia}
\affiliation{ARC Centre of Excellence in Low-Energy Electronics Technologies, UNSW Node, The University of New South Wales, Sydney 2052, Australia}
\affiliation{Dipartimento di Fisica "E. R. Caianiello", Universit\`a di Salerno, IT-84084 Fisciano (SA), Italy}
\author{Shanshan Liu}
\affiliation{Department of Chemistry, National University of Singapore, Singapore 117543, Singapore}
\author{Kian Ping Loh}
\affiliation{Department of Chemistry, National University of Singapore, Singapore 117543, Singapore}
\author{Dimitrie Culcer}
\affiliation{School of Physics, The University of New South Wales, Sydney 2052, Australia}
\affiliation{ARC Centre of Excellence in Low-Energy Electronics Technologies, UNSW Node, The University of New South Wales, Sydney 2052, Australia}

\begin{abstract}
Recent experiments have reported nonlinear signals in topological materials up to room temperature. Here we show that this response stems from extrinsic spin-orbit contributions to \textit{both} impurity and phonon scattering. While skew scattering dominates at low temperatures, the side jump contribution $\propto \tau/\tau_\beta$, where $\tau$, $\tau_\beta$ are the momentum and skew scattering times respectively. Consequently side jump exhibits a weak temperature dependence and remains sizable at room temperature. Our results provide a roadmap for engineering nonlinear transport at ambient conditions.
\end{abstract}

\maketitle

\section{Introduction}

\textit{Introduction}. Nonlinear electrical transport has emerged as a powerful probe of symmetry \cite{HeZhihai2021, HuangMeizhen2022, DuanJunxi2022, Tien2025, Okyay2025}, 
band geometry \cite{Kaplan2024,MercaldoMaria2025}, and scattering processes in quantum materials \cite{Burgos2023}. A striking manifestation is the nonlinear Hall effect (NLHE), which occurs even in time-reversal symmetric systems ~\cite{SodemannPRL2015, SNandy2019}. Remarkably, recent experiments have reported a robust NLHE persisting up to room temperature in spin-orbit-coupled materials, notably in bismuth~\cite{Makushko2024}, BaMnSb$_2$~\cite{Min2023}, and Sb-doped topological insulators such as MnBi$_4$Te$_7$~\cite{WangShaoyu2024}, among others \cite{Nishijima2023, HanJiahao2024, Min2024}. This offers significant potential for practical applications, 
for instance in terahertz technologies \cite{YangZhang2021, Guo2022, HuZhen2023, LiYun2024, Monch2025}, photo-detection \cite{WangHua2020,YulinShen2024}, 
non-volatile memory \cite{Xiao2020} and rectification \cite{Kumar2021, ChengBin2024, Kumar2024, LiuHao2025}.

Disorder is naturally present in realistic samples~\cite{SNandy2019, DuZZ2019, DuZZ2021NatureComm, DuZZ2021, OrtixCarmine2021}, and disorder-mediated contributions play a significant role even in $\mathcal{P}\mathcal{T}$-symmetric systems~\cite{Burgos2023, MaDa2023}. The Berry curvature dipole, often responsible for intrinsic responses~\cite{Zhou2020, ZhangCheng2022, ZhangChengLong2022}, is suppressed when additional crystal symmetries are present, as occurs, for instance, in topological insulator surface states with threefold rotational symmetry~\cite{SodemannPRL2015}. In such scenarios, extrinsic mechanisms -- side-jump and skew-scattering -- may dominate the nonlinear response~\cite{SNandy2019, KonigEJ2019, HirokiIsobe2020}. This was experimentally observed recently, with NLHE signals persisting up to room temperature \cite{ShanshanLiu2024}. The understanding of disorder in nonlinear responses remains limited, while the role of phonons, to our knowledge, has never been considered explicitly, precluding a systematic understanding of room-temperature effects. 
These observations highlight the need for a new theoretical perspective on disorder effects in nonlinear transport.

Motivated by this, our work focuses on nonlinear responses that arise from extrinsic scattering processes, particularly in regimes where intrinsic effects are suppressed by symmetry \cite{ShanshanLiu2024}. We develop a rigorous general framework based on the quantum Liouville equation that consistently incorporates disorder and phonon scattering as well as extrinsic spin-orbit coupling in the impurity and phonon scattering potentials. Our approach captures skew scattering and side jump stemming from both disorder and phonons, the temperature dependence of the nonlinear response, and the interplay between disorder and lattice effects. Our central result is the transversal nonlinear susceptibility 
\begin{equation}
\label{Eq:predictedequation}
\chi_{xyy}=C_1\frac{\tau^3}{\tau_{\beta}}+C_2\frac{\tau}{\tau_{\beta}},
\end{equation}
where $C_1$ and $C_2$ are functions of the Fermi energy, and $\tau_\beta$ is a characteristic skew scattering time associated with extrinsic spin-orbit scattering. Since the linear longitudinal conductivity $\sigma_{xx}$ is usually given by the Drude formula $\propto \tau$, the interplay of phonon and disorder scattering results in NLHE signals scaling as $\sigma_{xx}$ and $\sigma_{xx}^3$. The survival of the nonlinear signal up to room temperature \cite{ShanshanLiu2024} is explained as follows. When the temperature increases, both scalar scattering and spin-orbit scattering are affected, so that $\tau$ and $\tau_{\beta}$ decrease substantially, and the skew scattering contribution $\propto \tau^3/\tau_\beta$ becomes vanishingly small. However, the side-jump contribution $\propto \tau/\tau_\beta$ such that the temperature dependences of $\tau$ and $\tau_\beta$ partially compensate each other. This results in a weak temperature dependence of the side jump, so that the side jump signal remains strong at room temperature.

The results provide a key step toward understanding and controlling nonlinear charge dynamics for electronics and sensing. Firstly, the side-jump is expected to survive to high temperatures quite generally, making it a useful mechanism for room temperature applications. Secondly, given the strong mobility dependence of the skew scattering contribution, an improvement in the mobility by an order of magnitude could enhance the low-temperature nonlinear signal by 10$^3$. Finally, the strategy presented here can be extended to third order responses \cite{ShenLai2021, WangCong2022, SankarSoumya2024, NagTanay2023, GaoYang2023, HePan2024, ZhangXu2024, LiShengyao2024, MandarDebottam2024}.

\textit{General formalism}. Our approach is based on the solution of the quantum Liouville equation for the density matrix $\rho$, namely, $\partial_t \rho+ (i/\hbar)[H, \rho]=0$, where $H$ is the Hamiltonian of the systems. In general, the Hamiltonian is written as $H=H_0+H_E+U(\bm r)$, where $H_0$ is the unperturbed band Hamiltonian, $H_{E}(\bm r)=e\bm E\cdot \bm r$ is the external driving term and $U(\bm r)$ is the disorder potential. Importantly, in topological materials with strong spin-orbit interactions the scattering potential itself contains an extrinsic spin-orbit term. Such a term is known in topological insulators, where it affects STM quasiparticle interference~\cite{LeeWeiCheng2009} and weak localization~\cite{Adroguer2015}. In our case such an extrinsic spin-orbit term is present in the phonon as well as in the impurity scattering potentials.

The solution of the evolution equation for $\rho$ follows straightforward steps described in Ref.~\cite{Culcer2017, Burgos2022} (See also SM). 
Expressing the kinetic equation in the crystal momentum representation we find the equation
\begin{equation}
\label{Eq:generalsolutioncrystalmomentum}
\frac{\partial f}{\partial t}+\frac{i}{\hbar}[H_0, f ]+J_{0}(f)
=\frac{e\bm E}{\hbar}\cdot \frac{D f}{D \bm k}-J_{E}(f),    
\end{equation}
with $f_{\bm k}$ the disorder-averaged part of the density matrix. 
Also, the covariant derivative is defined as $\frac{D f}{D \bm k}=\frac{df}{d \bm k}+i[f,\bm{\mathcal{R}}]$ where the Berry connection vector reads
$\bm{\mathcal{R}}^{mm'}=i\langle u^{m}_{\bm k}| \nabla_{\bm k} u^{m'}_{\bm k}\rangle$ and $|u^{m'}_{\bm k}\rangle$ are periodic part of the Bloch eigen states of the unperturbed Hamiltonian. The collision integrals $J_{0}(f)$ and $J_{E}(f)$ are defined as
\begin{equation}
\arraycolsep 0.3ex
\begin{array}{rl}
\displaystyle J_0 (f) = & \displaystyle \frac{1}{2\pi\hbar}\int^{\infty}_{-\infty} 
d\epsilon 
\langle 
[U, G^{R}(\epsilon)[U,f]G^{A}(\epsilon)]
\rangle \\ [3ex]
\displaystyle J_E (f) = & \displaystyle \frac{1}{2\pi\hbar}\int^{\infty}_{-\infty} 
d\epsilon 
\langle
[U,G^{R}(\epsilon)[H_E, g_0 ]G^{A}(\epsilon) ] 
\rangle 
\end{array}
\end{equation}
with $G^{R/A}(\epsilon)$ retarded and advanced free Green's functions~\cite{Burgos2022}  and $\langle \cdots \rangle$ refers to the impurity average. Also, $g_0=\frac{1}{2\pi i}\int^{\infty}_{-\infty} 
d\epsilon G^{R}(\epsilon)[U, f]G^{A}(\epsilon)    
$ [See SM].
To account for the temperature dependence of transport properties, it is essential to include electron–phonon interactions, which yield an additional relaxation mechanism. The corresponding collision integral is given by 
\cite{Hwang2008,HuangBaoLing2008, ZhuXuetao2011, Kaasbjerg2012, Parente2013, Giraud2011, WangJie2013}:
\begin{widetext}
\begin{align}
\label{Eq:phononcollisionintegral}
[J_{ep}(f)]^{++}_{\bm k}
&=-\frac{2\pi}{\hbar}\sum_{\bm q}|D_{\bm q}|^2\delta(\hbar \omega_{\bm q}+\varepsilon^{+}_{\bm k-\bm q}-\varepsilon^{+}_{\bm k})
\left[N_{\bm q}f^{+}_{\bm k-\bm q} \left(1-f^{+}_{\bm k} \right)-(1+N_{\bm q})f^{+}_{\bm k}\left(1-f^{+}_{\bm k-\bm q} \right) \right] \nonumber \\
&-
\frac{2\pi}{\hbar}\sum_{\bm q}|D_{\bm q}|^2\delta(\hbar \omega_{\bm q}+\varepsilon^{+}_{\bm k}-\varepsilon^{+}_{\bm k+\bm q})
\left[\left(1+N_{\bm q} \right)f^{+}_{\bm k+\bm q}\left(1-f^{+}_{\bm k} \right)-N_{\bm q}f^{+}_{\bm k}\left(1 - f^{+}_{\bm k+\bm q}\right) \right], 
\end{align}
\end{widetext}
where $N_{\bm q}$  denotes the phonon occupation number. 
The electron–phonon coupling is described by the deformation potential $D_{\bm q}$. In the SM we provide the full derivation of the electron-phonon relaxation time.

\textit{Model Hamiltonian.}
The band Hamiltonian for topological insulators with warping can be cast as \cite{LiangFu2009, Naselli2022} 
\begin{equation}
\label{Eq:Hamiltonian}
H_{0}=\hbar v_F(k_x\sigma_y - k_y\sigma_x) + \frac{\lambda}{2}(k^3_{+} + k^3_{-})\sigma_z + \Delta\sigma_z,  
\end{equation}
with the definition $k_{\pm}=k_x\pm i k_y=ke^{\pm i\theta}$. The warping term can also be written as $H_{w}=\lambda k^3 \cos(3\theta_{\bm k})\sigma_z$
with $\theta_{\bm k}=\arctan(k_y/k_x)$. Written in this form, the warping term reflects the sixfold symmetry of the band Hamiltonian.  
The dispersion relation is
$\varepsilon^{\pm}_{\bm k} =\pm \sqrt{(\hbar v_Fk)^2+M^2_z}$, where $M_z = \Delta + \lambda k^3 \cos(3\theta_{\bm k})$. In order to have a non-linear response we need to break at least one mirror symmetry. In our systems this is done jointly by the gap and the warping parameter. 
For $\Delta=0$ the system has the sixfold symmetry while for $\Delta \ne 0$ the mirror symmetry is broken along $k_y$. The bigger the gap, the more the deviation from the sixfold rotation (See SM).

We consider here two independent sources of scattering. One is related the the scalar scattering potential $U_{0}(\bm r)$ that will define a relaxation time $\tau$. The other mechanism is related the extrinsic spin-obit coupling and can be cast as $U_{\bm k,\bm k'}=i\beta {\bm \sigma} \cdot \bm k \times \bm k'$, defining a relaxation time $1/\tau_{\beta}=\pi \rho(\epsilon_{F})[u^2_{0}\beta^2 k_F^4]/\hbar$, with $\rho(\epsilon_{F})$ the density of states at the Fermi energy and $\beta$ the extrinsic spin-orbit strength. 
Disorder plays an important role in the dynamics as experiments also show in a renormalization of the non-linear response \cite{HePan2021,ShanshanLiu2024,LuXiuFang2024}. This experimental fact motivates us to include two sources of scattering, where one will account for restoring $\mathcal{T}$-symmetry and one will account for stabilizing Fermi surface.

\textit{Leading order response.}
We solve the kinetic equation Eq.~\eqref{Eq:generalsolutioncrystalmomentum} for the distribution function $f_{\bm k}$ up to second order in the 
electric field \cite{Burgos2023}. For convenience in notation, we separate the function $f_{\bm k}$ into a band diagonal and an off-diagonal contribution as $f^{mm'}_{\bm k}=n^{mm'}_{\bm k}\delta_{mm'}+S^{mm'}_{\bm k}$. 
The leading distribution to second order in the electric field 
is diagonal in band index and satisfies the kinetic equation $[J_{0}(n^{(-2)}_{E2})]^{mm}_{\bm k} = (e/\hbar)\bm E\cdot  \nabla_{\bm k} 
n^{(-1)mm}_{E{\bm k}}$. Then we solve for $n^{(-2)mm}_{E2}$ in the relaxation time approximation. In the driving term, the distribution $n^{(-1)mm}_{E{\bm k}}$ is the leading order 
\textit{linear Boltzmann distribution}, which is straightforward to calculate \cite{Burgos2022}. Once we calculate the distribution $n^{(-2)}_{E2\bm k}$ we can use it to generate a 
new driving term. This follows from the collision integral of extrinsic spin-orbit coupling impurities. 
The kinetic equation to be solved is 
$[J_0(n^{(-2)}_{E2\beta})]^{mm}_{{\bm k}}=-[J_{\beta}(n^{(-2)}_{E2})]^{mm}_{{\bm k}}$, where the right hand side is a driving term. This will 
provide the nonlinear distribution $n^{(-2)}_{E2\beta} \propto \tau^3/\tau_{\beta}$, which will explain the cubic behavior of experimental data with 
respect to relaxation time [See SM for the full derivation]. After a straightforward calculation we find the susceptibility 
\begin{align}
\label{Eq:leadingresponse}
[\chi^{(-2)}_{xyy}]
&=
\lambda \left( \frac{\tau^3}{\tau_{\beta}} \right)
\left( \frac{e^3}{\hbar}  \right)
\frac{\rho(\varepsilon_{F})(1-\xi^2_{F})\varepsilon_F}{ \hbar^2 } 
H(\xi_{F}),
\end{align}
with the density of states $\rho(\varepsilon_{F})=\varepsilon_F/2\pi\hbar^2 v_F^2$ and dimensionless $H(\xi_F)=\left( 1/64 \right) \left(15\xi_{F}+13\xi^3_{F}-75\xi^5_{F}+33\xi^7_{F} \right)$. 

\begin{figure}[tbp] 
\centering
\resizebox{\columnwidth}{!} {    
\includegraphics[width=0.48\columnwidth]{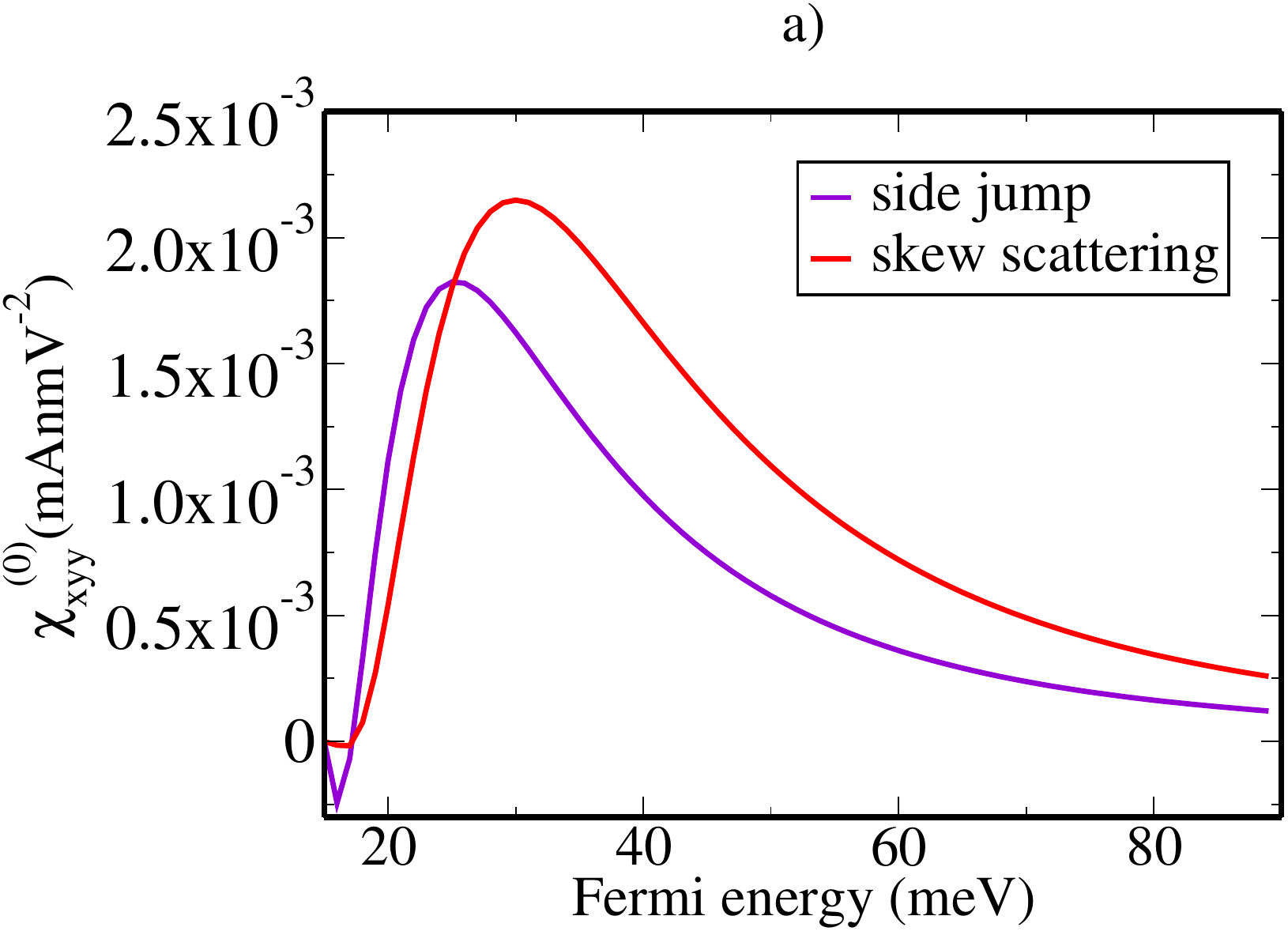}
\includegraphics[width=0.45\columnwidth]{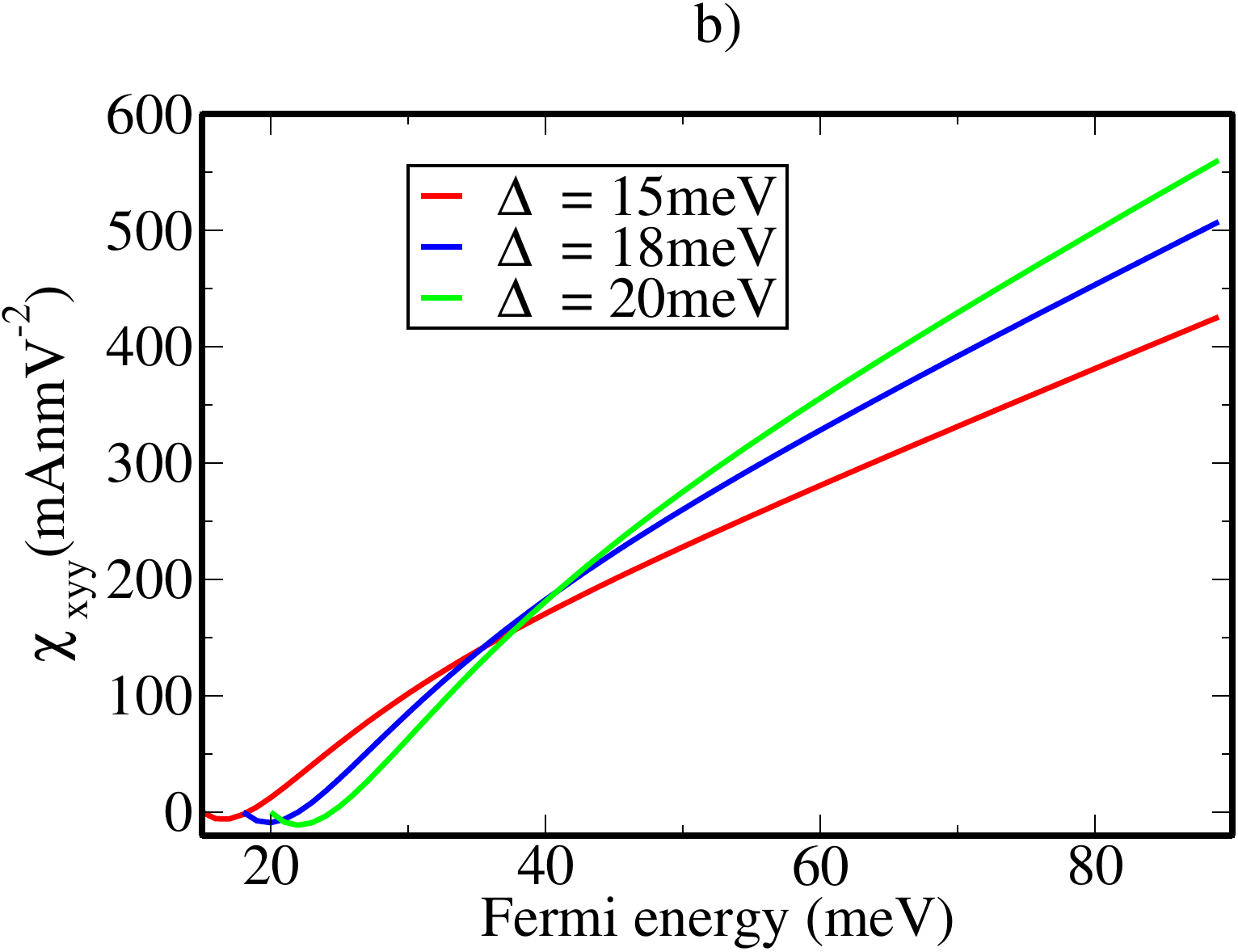}}
\caption{
a)Nonlinear sub leading susceptibility as a function of the Fermi energy. We take $v_F=1.6\times 10^{5}$m/s, the warping constant $\lambda=80$eV$\AA^3$ and $\tau=1$ps and $1/\gamma_{\beta}=0.5$ps. We fixed the $\Delta=15$meV.
b)Total nonlinear susceptibility as a function of the Fermi energy. We take $v_F=1.6\times 10^{5}$m/s, the warping constant $\lambda=80$eV$\AA^3$ and $\tau=1$ps and $1/\gamma_{\beta}=0.5$ps. 
}
\label{Fig:responseFermienergy}
\end{figure}

\textit{Sub-leading side jump effect.}
The nonlinear response observed in experiments also shows a linear behavior with the longitudinal linear conductivity which can be traced back to a linear in relaxation time behavior. This contribution is non trivial and represents our main result.  
In order to explain this linear in relaxation time behavior, the linear side jump effect is the main ingredient~\cite{Burgos2022}.
This means that semi classical coordinate shift~\cite{Sinitsyn2006,Sinitsyn_2007,Sinitsyn_2007JPCM}, although we calculate it here in a fully quantum mechanical fashion~\cite{Culcer2010,Burgos2022}, 
experienced by carriers migrates to the second order response to renormalise the response 
of $\mathcal{P}\mathcal{T}$-symmetric states above the Neel temperature.
The procedure is as follows: First we calculate a linear distribution $n^{(0)}_{E\bm k}$ independent of impurity density (but with disorder effects). 
It follows from the electric field corrected collision integral, namely, by solving the kinetic equation 
$[J_{0}(n^{(0)}_{E,\beta})]^{mm}_{\bm k} = - [J_{E,\beta}(n^{(0)}_{FD})]^{mm}_{\bm k}$. This provide 
the quantum mechanical version of the semi classical side jump effect~\cite{Burgos2022}.
With the knowledge of the linear distribution $n^{(0)mm}_{E,\beta}$ then we can solve the equation 
$[J_{0}(n^{(-1)}_{E2,\beta})]^{mm}_{\bm k} = (e/\hbar) \bm E \cdot \nabla_{\bm k} n^{(0)}_{E,\beta}$ in order
to calculate the nonlinear distribution $n^{(-1)}_{E2,\beta}$, which is already re-normalised by spin-orbit scattering potential. 
We also need to solve the kinetic equation  
$[J_{0}(n^{(0)}_{E})]^{mm}_{\bm k} = - [J_{E}(n^{(0)}_{FD})]^{mm}_{\bm k}$ to calculate the distribution 
$n^{(0)mm}_{E}$, which is again a linear side jump effect. 
Using once more the kinetic equation 
$[J_{0}(n^{(-1)}_{E2})]^{mm}_{\bm k} = (e/\hbar) \bm E \cdot \nabla_{\bm k} n^{(0)}_{E}$ we get the diagonal nonlinear
distribution $n^{(-1)mm}_{E2}$, which is not re-normalised by the extrinsic spin-orbit scattering. Both $n^{(-1)mm}_{E2,\beta}$ and $n^{(-1)mm}_{E2}$ scales 
linear in relaxation time $\tau $ and can be used now to generate the off-diagonal channels which are re-normalised 
by the spin-orbit scattering potential and are effectively zeroth order in impurity density. 
They follow by solving the kinetic equation
$S^{(0)mm'}_{E2\bm k,\beta}=i\hbar (\epsilon^{m}_{\bm k}-\epsilon^{m'}_{\bm k})^{-1}([J_{0}(n^{(-1)}_{E2,\beta})]^{mm'}_{\bm k}+
[J_{\beta}(n^{(-1)}_{E2})]^{mm'}_{\bm k})$[See SM for details].
The current follows by tracing the off-diagonal velocity $v^{mm'}_i=i\hbar^{-1}(\varepsilon^{m}_{\bm k} - \varepsilon^{m'}_{\bm k})\mathcal{R}^{mm'}_{i}$ with the off-diagonal distribution, which formally scales linear in relaxation time.

\begin{figure}[tbp] 
\centering
\resizebox{\columnwidth}{!} {    
\includegraphics[width=0.49\columnwidth]{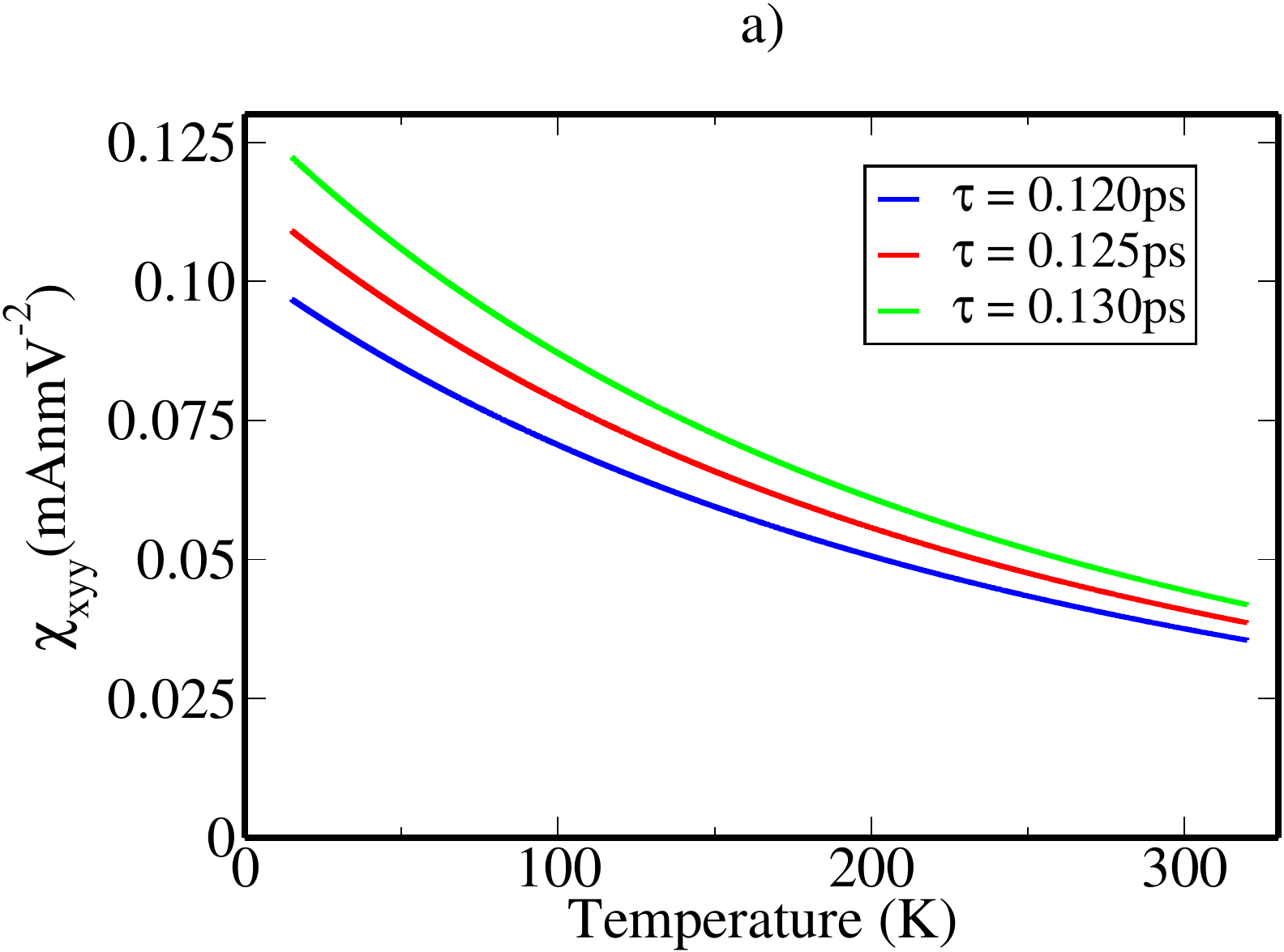}
\includegraphics[width=0.47\columnwidth]{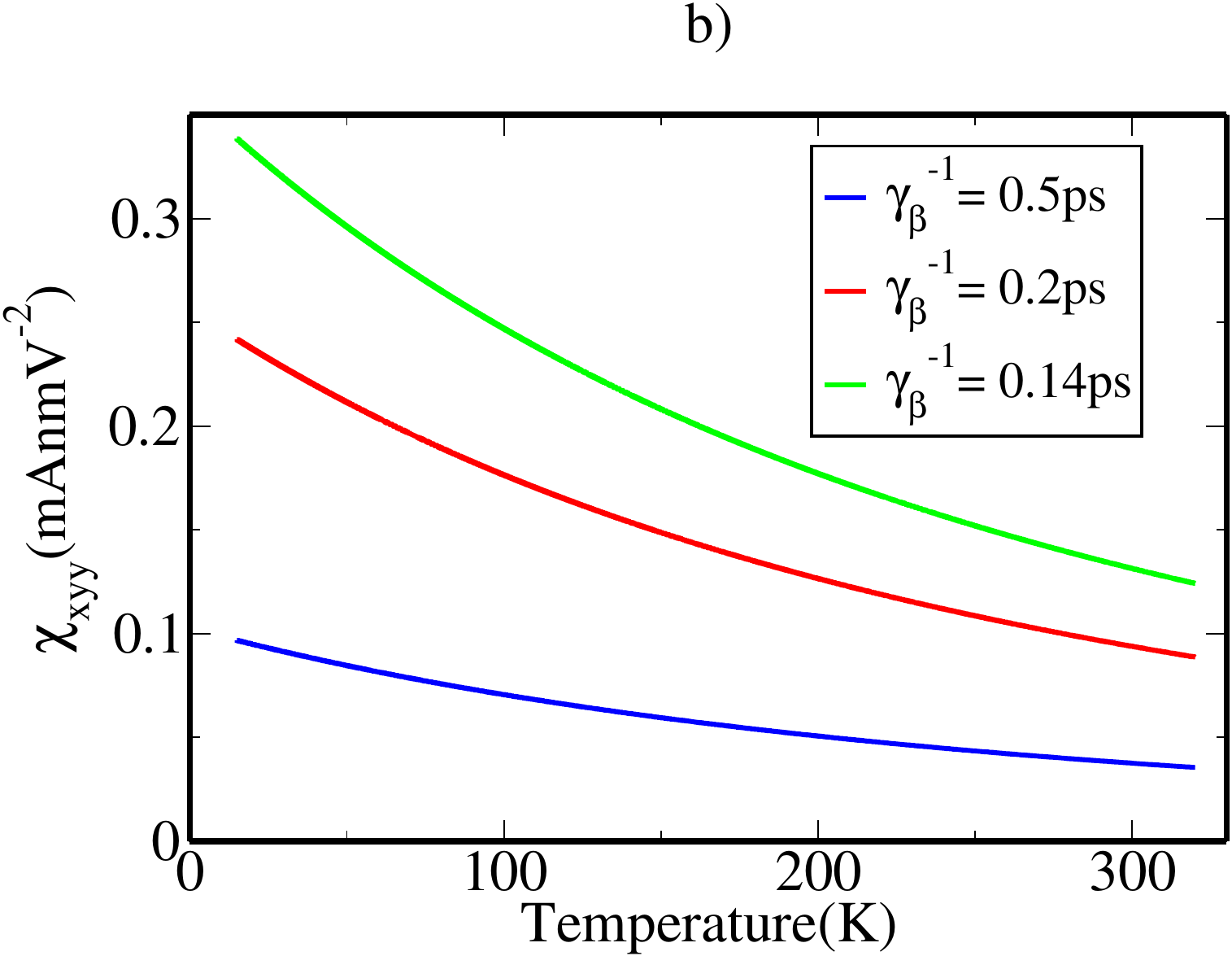}}
\caption{
Total nonlinear susceptibility as a function of the temperature. 
We take $v_F=1.6\times 10^{5}$m/s, the warping constant $\lambda=80$eV$\AA^3$ and $\Delta=15$meV, $\varepsilon_F=25$meV 
a)and $\gamma_{\beta}=0.2$ps and changed the momentum relaxation time 
b) we fixed $\tau=0.12$ps and changed $\gamma_{\beta}$. 
}
\label{Fig:responseT}
\end{figure}

\textit{Sub-leading skew scattering effect.}
The procedure just described will provide the sub-leading response related to non-linear side jump like effect. As is well known, 
there is a skew scattering effect of the same order of the side jump contribution \cite{Burgos2022,Burgos2023}. Such an effect arises
when we use the off-diagonal distributions to generate new driving terms.
It follows by solving the kinetic equation $[J_0(n^{(0)}_{E2,\beta})]^{mm}_{\bm k} = -[J_0(S^{(0)}_{E2,\beta})]^{mm}_{\bm k}  -  [J_{\beta}(S^{(0)}_{E2})]^{mm}_{\bm k}$. The renormalised off-diagonal distribution is already known, however
note that the second driving term requires the $\beta$-independent off-diagonal distribution 
$S^{(0)mm'}_{E2\bm k}=i\hbar (\epsilon^{m}_{\bm k}-\epsilon^{m'}_{\bm k})^{-1}([J_{0}(n^{(-1)}_{E2,\beta})]^{mm'}_{\bm k})$.
The renormalised side jump and skew scattering distributions will explain the experimental linear in relaxation time behavior of the 
nonlinear response [See SM for more details]. The full sub-leading susceptibility including side jump and skew scattering is written as
\begin{align}
\label{Eq:subleadingresponse}
[\chi^{(0)}_{xyy}]
&=-
\lambda 
\left( \frac{\tau}{\tau_{\beta}} \right)
\left( \frac{e^3}{\hbar} \right)
\frac{\rho(\varepsilon_{F})}{\epsilon_{F}} 
(1-\xi^2_{F} )G(\xi_F)
\end{align}
with the dimensionless function 
\begin{equation}
\begin{split}
&G(\xi_F)=
\frac{4}{512}
\Bigg\{
(1-\xi^2_{F})
[(10\xi^3_{F}+54\xi^5_F-36\xi^7_F)]  \\ 
&\quad
+
(1 - \xi^2_{F} )^{1/2} \left[-88\xi^3_{F} + 128\xi^5_{F} - 24\xi^7_F \right]
\Bigg\}. 
\end{split}
\end{equation} 

We plot the separate contributions in Fig.~\ref{Fig:responseFermienergy} a). 
It should be noted that side jump and skew scattering can not be separated in experiments: they are both different functions of the Fermi energy
but of the same order of magnitude, as is the case in linear response~\cite{Burgos2022}. Note that the function $G(\xi_F)$ scales 
with powers of inverse Fermi energy, what explains the rapid decay of the sub-leading contribution.

Note that in the diffusive regime considered here the susceptibility $\chi^{(-2)}_{xyy}$ dominates, which is explicit 
in the prefactor $\varepsilon_F\tau/\hbar \gg 1$ in Eq.~\eqref{Eq:leadingresponse}. We plot 
the $\chi_{xyy}=\chi^{(0)}_{xyy}+\chi^{(-2)}_{xyy}$ in Fig.~\ref{Fig:responseFermienergy}b) as a function of the Fermi energy.
The increasing behavior with Fermi energy is also captured by experiments~\cite{ShanshanLiu2024}. Such a behavior is dominated by $\chi^{(-2)}_{xyy}$ 
precisely by the factor indicating diffusive regime which scales as $\sim \varepsilon^2_F$. 
Note that the dominant contribution in $H(\varepsilon_F)$ scales as $\sim 1/\varepsilon_F$. All this 
will provide the linear increasing of the response with the Fermi energy.

Scattering by phonons becomes relevant as the temperature increases. Solving the electron-phonon collision integral (see SM) in the relaxation time approximation
\begin{align}
\frac{1}{\tau_{ph}}
&\approx
\left(\frac{k_BT}{\hbar  }\right) 
\left( \frac{c^4_{R}C}{2c^4_l\pi\eta \rho_{M} c^2_R } \right)  
\left(\frac{\alpha^2}{\hbar^2} \right) (2k_F).
\end{align}
where we have used the deformation potential of Ref.~\cite{Parente2013}:
\begin{equation}
D_{\bm q}=
\alpha \sqrt{\frac{C}{qA}} \left( \frac{\omega^{(0)}_{q}}{c_l} \right)^2\sqrt{\frac{\hbar}{2\rho_{M}\omega^{(0)}_q }} \Lambda_{l}(q),
\end{equation}
with $\Lambda_{l}(q) \approx 1$ and $A$ is the area of the unit cell. For Bi$_2$Te$_3$ the parameter $\alpha$=35eV, $C=1.2$, $c_l=2800$m/s is the longitudinal phonon velocity. $\rho_M$=7861kg/m$^3$ is the mass density of the material and $\omega=c_Rq$ where $c_R=0.86c_t$ with the transversal phonon velocity $c_t=1600$m/s. We recall that $1/\eta = M/v^2_F$ is an effective mass, where $M$ is the gap (see SM).

\textit{Discussion.}
The BCD has attracted interest in $\mathcal{P}\mathcal{T}$-broken systems \cite{MaQiong2019,Sinha2022} due to its sensitivity to band structure~\cite{YangZhang2018, DuZZ2018,FacioJorge2018, ZhangYang2018}, although crystal symmetries restrict its presence. Even when allowed, side-jump and skew-scattering contributions can compete with the BCD at the same order, all scaling linearly with $\tau$, making experimental distinction difficult.
In contrast, for $\mathcal{P}\mathcal{T}$-symmetric systems—where BC is forbidden—the quantum geometric tensor (QGT) has emerged as a key intrinsic mechanism \cite{ChongWangPRL2021, HuiyingLiu2021, AnyuanGao2023, DasKamal2023}. These responses follow a specific $\tau$-scaling \cite{Burgos2023}: a Drude term scaling as $\tau^2$ and intrinsic terms scaling as $\tau^0$. Importantly, $\tau^0$ denotes mathematical independence from impurity density, arising from competing relaxation processes, both scale linear with impurity concentration, a phenomenon also seen in linear response \cite{Sinitsyn_2007JPCM,Burgos2022}.

At finite temperatures, the disorder dependence changes due to two effects: (i) Above the N\'eel temperature, thermal fluctuations restore time-reversal symmetry \cite{WangNaizhou2023}, placing the system in a $\mathcal{P}\mathcal{T}$-broken regime and modifying its scaling. Experiments have reported $\tau^3$ and $\tau$ contributions \cite{Kang2019, HePan2021, Tiwari2021, HePan2022}, explained via a phenomenological model inspired by the anomalous Hall effect \cite{Kotzler2005, Nagaosa2010}. 
(ii) Phonon-induced relaxation becomes important. To account for this, we apply Matthiessen’s rule: $\tau^{-1} \rightarrow \tau^{-1} + \tau^{-1}_{ph}$ and similarly for $\tau^{-1}_{\beta}$. The parameter $\tau_{\beta}$ captures an effective relaxation channel that renormalizes the nonlinear conductivity.

\begin{figure}[tbp]
\centering
\includegraphics[width=0.4\textwidth]{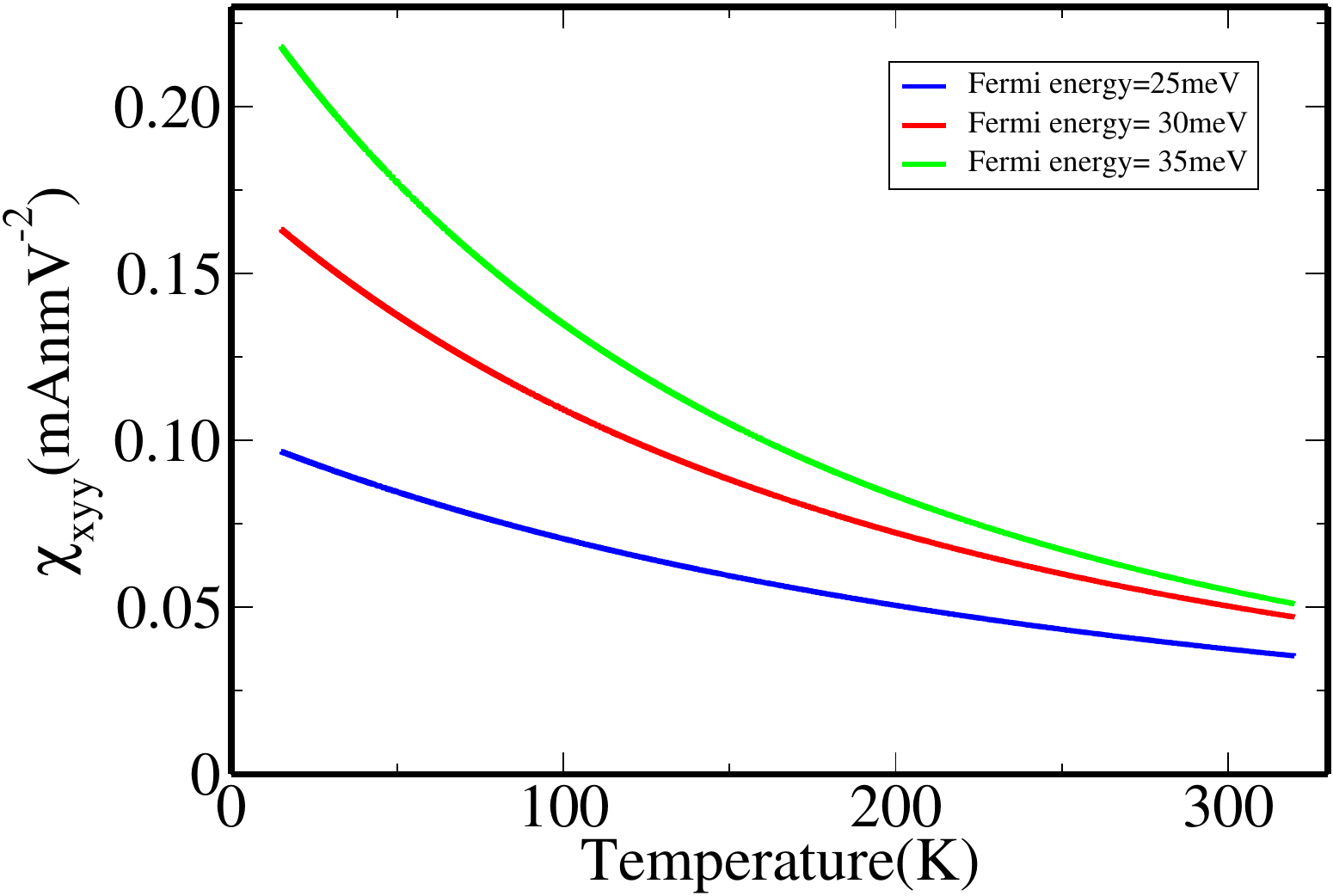}
\caption{Total nonlinear susceptibility as a function of the temperature. 
We take $v_F=1.6\times 10^{5}$m/s, the warping constant $\lambda=80$eV$\AA^3$, $\tau=0.12$ps,
$\tau_{\beta}=0.5$ps and $\Delta=15$meV.}
\label{Fig:responseTsecond}
\end{figure}

Applying Matthiessen’s rule to Eq.~\eqref{Eq:predictedequation}, the second-order susceptibility takes the form $\chi_{xyy}(T) = (1 + AT)[B(1 + DT)^{-1} + F(1 + DT)^{-3}]$, where $A$ and $D$ encode the relative strength of phonon scattering versus spin-orbit and scalar disorder, respectively. The prefactors $B$ and $F$ depend on microscopic parameters such as the Fermi energy and gap. Using Eqs.~\eqref{Eq:leadingresponse} and \eqref{Eq:subleadingresponse}, Fig.~\eqref{Fig:responseT} shows $\chi_{xyy}(T)$ and its sensitivity to $\tau$ and $\tau_{\beta}$, with the latter having stronger impact as it scales both contributions. In Fig.~\eqref{Fig:responseTsecond}, we plot $\chi_{xyy}(T)$ varying the Fermi energy while keeping the relaxation times fixed. This behavior, observed in e.g. Ref.~\cite{LuXiuFang2024}, suggests skew scattering ($\sim \tau^3$) dominates. The sub-leading channel arises from a \textit{linear side jump effect} extended to second order, scaling as $\sim \tau$. These contributions — nonlinear side jump and skew scattering — have not been previously computed but become relevant at low Fermi energies, where interband coherence enhances quantum effects \cite{GholizadehSina2023, NakazawaKazuki2025}. Our framework clarifies these mechanisms \cite{SuarezRodriguez2024, CaoZheng2025} and exposes the shortcomings of approximations that neglect their interplay.

\textit{Comparison to experiments.}
To analyze the way the nonlinear Hall response scales with disorder, a phenomenological scaling relation was proposed in Ref.~\cite{DuZZ2019} under the assumption of time-reversal symmetry. In that framework, linear scaling of the nonlinear Hall conductivity with the linear longitudinal conductivity was attributed to contributions from the Berry curvature dipole (often labeled intrinsic), side-jump, and skew scattering mechanisms \cite{YouJhihShih2018, Shao2020}. This approach was later adopted in experiments on the surface states of topological insulators \cite{HePan2021}, where a Berry curvature triple (BCT) was introduced as the driving mechanism behind the skew scattering—assumed to be constant—and associated with an overall scaling of order $\tau^3$. Such an interpretation is consistent with systems that explicitly break $\mathcal{P}\mathcal{T}$ symmetry. By contrast, the $\mathcal{P}\mathcal{T}$-symmetric state studied in this work presents a fundamentally different scenario. In this case, the Berry curvature vanishes identically at each $\bm{k}$ due to symmetry, making the definition of a BCT ambiguous or even meaningless. Additionally, the quantum metric dipole is zero, eliminating all intrinsic second-order contributions. This leaves disorder-induced mechanisms as the sole source of nonlinear response. Our main result, Eq.\eqref{Eq:predictedequation}, differs from the analysis in Ref.\cite{HePan2021} in a crucial way: here, the skew scattering—represented by the relaxation time $\tau_\beta$—is not treated as a fixed or frozen quantity. Instead, both $\tau$ (the scalar relaxation time) and $\tau_\beta$ are governed by the same impurity concentration. This refinement is supported by recent experiments \cite{ShanshanLiu2024}, which demonstrate that phonon interactions can influence skew scattering, aligning well with our theoretical predictions.

\textit{Conclusion}. We have studied the impact of disorder and phonon scattering on the nonlinear response of gapped topological insulators with hexagonal warping. Our density matrix formalism provides a consistent description of room-temperature experimental results. Crucially, it shows that both skew scattering and side-jump mechanisms make relevant contributions, even at subleading order. These results underscore the need for a careful treatment of relaxation processes beyond simplified models, and provide a versatile framework for other spin-orbit coupled systems and disorder scenarios.


\bibliographystyle{apsrev4-1}
\bibliography{WarpinTI}

\end{document}